\documentclass[review]{elsarticle}

\usepackage{hyperref,}
\usepackage[cmex10]{amsmath}

\journal{Energy}

\begin{document}

\begin{frontmatter}

\title{Extreme prices in electricity balancing markets from an approach of statistical physics}

\author{Mario Mureddu and Hildegard Meyer-Ortmanns}
\address{Physics and Earth Sciences\\
	Jacobs University Bremen\\
	28759 Bremen, Germany\\
	Email: h.ortmanns@jacobs-university.de}

%
%

\begin{abstract}
An increase in energy production from renewable energy sources is viewed as a crucial achievement in most industrialized countries.
 The higher variability of power production via renewables leads to a rise in ancillary service costs over the power system, in particular costs within the electricity balancing markets, mainly due to an increased number of extreme price spikes.
This study focuses on forecasting the behavior of price and volumes of the Italian balancing market in the presence of an increased share of renewable energy sources. Starting from  configurations of load and power production, which guarantee a stable performance, we implement fluctuations in the load and in renewables; in particular we artificially increase the contribution of renewables as compared to conventional power sources to cover the total load. We then forecast the amount of provided energy in the balancing market and its fluctuations, which are induced by production and consumption. Within an approach of agent based modeling we estimate the resulting energy prices and costs. While their average values turn out to be only slightly affected by an increased contribution from renewables, the probability for extreme price events is shown to increase along with undesired peaks in the costs.
\end{abstract}


\end{frontmatter}


\section{Introduction}

The increasing environmental awareness, together with the progressive reduction of production and installation costs\cite{Goyena2009}, leads to a considerable growth in the amount of Renewable Energy Sources (RES) that is installed worldwide.
Moreover, the increasing propensity to reduce greenhouse gas emissions requires an increment of the energy produced by clean, accessible energy sources such as wind and photovoltaic (PV) generation.
Despite the great advantages of these energy sources, their intrinsic variability in power production badly fits to the very hierarchical structure and the strictly dispatch rules of actual power systems. The limited accuracy of the prediction of their energy production profiles makes the management of these intermittent power sources difficult and limits the amount of RES generation that the power system can tolerate.

After the network liberalization over the last 15 years, the system balancing in real time is performed via the Electricity Balancing Market (EBM), which is a subphase of the Ancillary Services Market (ASM). This market phase shall ensure the correct balanced state over the system at the transmission level, providing the security of the supply at the lowest possible costs. However, the short time-scale and the volatility of this market phase produce higher energy costs when compared with the day-ahead market phase. Therefore an increase in the EBM volume can lead to very high system  maintenance costs.
The growing amount of the installed RES generation introduces a high number of partially correlated fluctuations in the power production. Along with that, it becomes  more difficult to predict the amount of energy that is needed for balancing the system.

In general, an increase in  production fluctuations could lead to both an increase in market average volumes and a more frequent occurrence of extremely high values of the volume. Whereas an increase in average volumes could be cured by strengthening the reserve capacity, the occurrence of extreme volumes is more difficult to control. Moreover, given the fact that the relation between price and demand, also known as power stack function \cite{eydeland1999fundamentals, geman2006understanding}, is highly nonlinear, large volume events can lead to very high energy prices. Such extreme and unwanted price events have been  observed already by Nicolosi in \cite{Nicolosi2010} for the German system: if they happen too often, the total costs of the market session increase and undermine the principles on which the electricity market was designed. Therefore, the forecast of the fluctuations' impact on the balancing market can be vital for an optimal planning of the network growth, and for uncovering possible critical situations of the network.  So it is not surprising that the evaluation of volumes and prices of EBMs, and in general, of electricity markets, has attracted  much  interest in the last years. However, the proposed solutions are mostly based on historical data for the market volume together with learning procedures of agents and game theory \cite{nogales2002forecasting, contreras2003arima}, \cite{weidlich2008critical, Tellidou2007, Xiong2002}. They need an update to more recent data and an extrapolation towards future increased contributions from RES.

In the next section we shall show  how a combination of an approach from statistical physics with agent based modeling overcomes the need for using historical data and allows for the prediction of energy prices, the emergence of price peaks, volumes of balancing markets and overall daily costs for different contributions from RES, to cover  up to 60\% of the load. This combination of methods was proposed and validated for the Italian EBM in \cite{MuredduPONE2015}. Our results together with the methodology should enter planning procedures of how to  further increase and control the amount of RES in the future.

\section{Methods} 	\label{sec:methods}
Before going into detail, let us summarize the procedure, which consists of three steps: (i) Based on real data for production and consumption at a certain representative day in the winter period of 2011-2012 in the Italian grid, we generate a certain set of starting configurations, each one describing a combination of production and load at nodes of the Italian transmission grid, which lead to a stable performance by construction. The real data were taken every 15 minutes over a whole day for all 6 price zones of the power grid in Italy. We then extrapolate these data towards a higher contribution of RES, ranging from the real value of 24\% to 60\% in the extrapolates sets. In all extrapolated cases we guarantee a stable performance by running optimal DC-power flow equations to adjust the production by conventional generators so as to guarantee an overall balanced power in the grid.

(ii) Each of the configurations of the resulting set (6 zones x 96 time instants per day for 24, 30, 40, 50, and 60\% of RES) serves as starting point for generating an ensemble of configurations in the spirit of statistical physics. In statistical physics one usually describes a macrostate say of a gas of molecules by an ensemble of microstates; microstates differ by small deviations of the generalized coordinates and momenta, so that an average over many microstates leads to representative macroscopic observables. In analogy here, members of each ensemble differ from the starting configuration and therefore also mutually by small deviations in the load and the renewable energy production, chosen from a Gaussian or Weibull distribution, as explained below. Mean and width of the Gaussian distribution as well as the parameters of the Weibull distribution are chosen from real data and depend on the load and the type of renewable energy. So each configuration j of the ensemble represents a certain realization of fluctuations in load and production, for which quantities like the resulting  mismatch $S_i^j$ of power at node i, induced by the various fluctuations, can be measured. $S_i^j$ can be summed over all nodes of the grid to obtain the total mismatch $S^j$ in power production for a given configuration j, which the balancing market is supposed to compensate for. To obtain  representative values of the mismatch (later called the market volume), also here (as for the microstates of a gas) a sufficient number of configurations should be included in the ensemble.

(iii) The energy balancing market
The energy balancing market is modeled by a so-called market authority and a set of agents. The market authority knows the required amount of power $S^j$, which is needed for balancing consumption and production; it accepts or rejects  bids from the agents until an amount of $S^j$ is obtained at the lowest possible price. It then informs the agents about the outcome of their placed bids as to whether they were accepted or not. The agents are assigned to conventional generators in a one-to-one relation (for simplicity). They choose their bids from a distribution of propensities to offer a certain amount of energy at a certain price. The propensity distribution changes with time during the learning phase, using a modified Roth-Erev algorithm\cite{Nicolaisen2001}. For the learning phase we choose 3000 updates of the propensity distribution, in which the agents are trained on the same number of different configurations, chosen from the ensemble around a fixed starting configuration, so differing just by fluctuations among each other. This number of updates turned out to be sufficient for the propensity distribution to converge towards an optimized distribution, resulting from the learning experience of the agents after feedback from the market authority.

For the next thousand configurations of a given ensemble the propensity distribution of agents is then kept fixed, and the energy price in this market session can be calculated,  for each configuration, leading to a distribution of energy prices over all configurations of the ensemble. The distribution of energy prices refers to a certain time during a day for a given zone in Italy. Repeating the whole procedure for different starting configurations, corresponding to different instants of time at the reference day and different price zones, we can measure histograms of how often a price from a certain price interval was achieved over the day or for a restricted time interval of an hour etc.. We are particularly interested in the shape of these histograms as a function of the percentage of renewables, which contributed to the power production. Details are presented in the following sections.

\subsection{Evaluation of imbalances of real-time systems} \label{sec:perturbation}
Let us first estimate the effect of RES and load power fluctuations on the system's power balance.
According to the literature\cite{Castronuovo2004, Pappala2009, Lu2013}, wind, PV and load forecasting errors in the power production are often treated as normal-distributed. So their power production or consumption can be modeled in a statistical way, assuming truncated Gaussian-like forecast errors with standard deviations $\sigma_i$, where the errors represent the expected power variations at each single fluctuating element $i$ of the power grid at a given time.
The associated variables are the following:
\begin{itemize}
	\item load power demand $D_l$, the corresponding  standard deviation $\sigma _l$ of the forecasting error and minimum and maximum values of the distribution, $m_l$ and $M_l$, respectively, corresponding to the load-power constraints;
	\item wind power production $G_w$ and the corresponding $\sigma _w$, together with the  power constraints $m_w$ and $M_w$ of the generators;
	\item photovoltaic power production $G_{PV}$, the corresponding $\sigma _{PV}$, and the production limits $m_{PV}$ and $M_{PV}$, corresponding to the  power constraints of the PV-generators.
\end{itemize}
According to these constraints, a possible state of the system can be  sampled numerically by adding a random value to the expected power production and consumption at every node and RES generator $i$ of the grid. The random variable is  extracted from the truncated normal distribution, whose probability density function (PDF) is defined in equations \ref{eq:normal_def} and \ref{eq:normal_pdf}.
\begin{equation} \label{eq:normal_def}
N_{PDF}^T =
\left\{ \begin{array}{rl}
0 &\mbox{ if $x<m_i$} \\
N_{PDF} &\mbox{ if $m_i<x<M_i$} \\
0 &\mbox{ if $x>M_i$}
\end{array} \right.
\end{equation}

\begin{equation} \label{eq:normal_pdf}
N_{PDF}(x) = \frac{1}{\sqrt{2\pi\sigma_i^2}} e^{\frac{x^2}{2\sigma_i^2}}.
\end{equation}

The outcome of this procedure is one of the possible configurations or states $j$, in which the system in zone k ($k=1,...,6$) can be found in real time, due to assumed unavoidable fluctuations of RES and load. In order to check the impact of other than Gaussian-type fluctuations, we complemented the normal distribution of (\ref{eq:normal_pdf}) in case of wind production  by a Weibull distribution, whose PDF is defined in equation \ref{eq:weibull}, while the fluctuations for photovoltaics production and load were kept being chosen from Gaussian distributions. Since the literature gives values between 1.5 and 3 for the value of the parameter $a$ \cite{Lun2000, Seguro2000}, we have chosen $a = 2$ and $\lambda = \frac{P_w}{\Gamma(\frac{3}{2})}$, where $P_w$ here is chosen as the wind production from the reference configuration and $\Gamma$ denotes the Gamma-function. The PDF is then given as
\begin{equation} \label{eq:weibull}
f(x; \lambda, a ) =
\begin{cases}
\frac{a}{\lambda} (\frac{x}{\lambda}) ^{a - 1}  e^{-( \frac{x}{\lambda} )^{a}},& \text{if } x\geq 0\\
0,              & \text{otherwise}.
\end{cases}
\end{equation}

Starting from the so generated configuration or network state, we apply the optimal DC-power flow algorithm \cite{Zimmerman2011} to calculate the power $S^j$ that is needed for balancing the mismatch in power, induced by the deviations from the starting configuration, for each state $j$ of the ensemble, related to zone k and to each of the 96 time instants a day. Due to the stochastic nature of RES and load fluctuations, also $S^j$ is a random variable.
Therefore, to sample sufficient statistics of the market behavior, a significant number of possible balancing requirements is needed. It is obtained by numerical sampling a large number of possible perturbed configurations $j$, each one with an associated balancing requirement $S^j$. Its distribution over the ensemble and over the day  is then  used for describing the daily expected volume of the balancing market in the system.

\subsection{Agent Based Modeling} \label{sec:abm}
Energy prices and total costs in the EBM are determined by an agent based modeling approach, for which we use a  modified Roth-Erev algorithm, introduced by Nicolaisen et al. \cite{Nicolaisen2001} in 2001 and used by Rastegar et al.\cite{Rastegar2009} already for the simulation of the Italian ODA electricity market. The electricity-market operators are represented by agents, who learn how to place optimal bids in competitive auctions with the aim of buying (or selling) in the most profitable way. In order to simulate how real market operators acquire knowledge about the market in the course of time and adapt their decisions,
Roth-Erev algorithms simulate this learning process by adjusting  the offer propensities of agents in a self-consistent way with the goal to  maximize profits. Market operators pursue economic guidelines, when they represent power plants (or groups of them) in the EBM auction phase. They are allowed to place bids into the EBM auction, in which they must specify how much the corresponding power plants can vary their amount of power supplied to the system, and at what price they will offer this service. For simplicity we represent each conventional power-plant generator by a single agent, although the exact relationship among market operators and brokers may vary over time and can be more involved.

Each agent $k$ is allowed to offer an amount of power $g^k_{off}$ that must meet the physical constraints of the power generator k:
\begin{itemize}
	\item $G^k_{min} \leq G^k_{given} + g^k_{off} \leq G^k_{max}$, where $G^k_{min}$ and  $G^k_{max}$ are the minimum and maximum power production constraints of the generator, respectively, and $G^k_{given}$ is its actual power production.
	\item $-G^k_{ramp} \leq g^k_{off} \leq G^k_{ramp}$, where $G_{ramp}$ is the generator ramping constraint. (Depending on the technology, each generator has ramping constraints, which limit its maximum change in power production $G_{ramp}$ in time.)
\end{itemize}
In order to define the bids' price, we use the concept of agent propensities, representing the willingness of each agent to place a bid at a certain price on the market.  The offer propensities of each operator $k$  are described in terms of a discrete set of  probabilities, $q^i_k$, to be defined below, corresponding to possible  bidding strategies $\{(m^i_k, s^i_k)\}$, which roughly speaking differ by how to deal with risks in offering  higher prices.
The index i, $0<i<N$, labels the strategy, $N$ is the number of possible strategies,  and $s^i_k$ is the k-th operator's  propensity to make an offer at a given (so-called markup) value $m^i_k$  ($1\leq m^i_k \leq 10$ for upward bids, $0\leq m^i_k \leq 1$ for downward bids). The number of strategies equals the number of intervals into which the range of $m^i_k$ is divided. Here we have chosen $N=50$, so that one has to assign 50 propensities to values of $m^i_k$. The markup value determines  the bidding price according to  $p_{off}^k = C_{prod}^k\cdot m^\star_k$, where $C_{prod}^k$ is the production cost (per MWh) of each generator k, given by its technology type, labeled by the subscript $prod$, and $m^\star_k$ is the actual chosen value from the discrete distribution for the bid of agent k. So the operators' behavior is modeled stochastically, where the probability of placing a bid at a given price $p_{off}^k = C_{prod}^k\cdot m^\star_k$ is given by the normalized propensity $q^i_k = s^i_k/\sum_i s^i_k$ with $i=1,...,N$ and $k=1,...,G$ with G the number of conventional producers equal to the number of agents. It is then the set of propensities, which get optimized when updated in an iterative reinforced learning algorithm.  Initially, all propensities $s^i_k$ are set to the same value $s^i_k=1$. Each learning iteration step is divided into three phases:

\begin{enumerate}
	\item Bid presentation: Every agent k presents a bid $\left(g_{off}^k,p_{off}^k\right)$, both for the upward and downward market. This bid is given by a feasible quantity of offered energy $g_{off}^k$ (i.e. satisfying the physical constraints) and by a price $p_{off}^k$, which will be drawn from the agents' propensities.
	\item Market session: Given the knowledge of the total balancing needs of the system $S^j$, all the bids, which are needed to ensure sufficient energy supply, are checked with respect to their economic profit and  the physical constraints of the system.
	\item Agent update: Market outcomes are communicated by the market authority to each agent, who updates his propensities in relation to the profit, which he made in the previous session. The agents' propensities at iteration step $t$ are updated as follows:
	\begin{equation}
	s^i_k(t)=(1-r)\cdot s^i_k(t-1) + E_i(t),
	\end{equation}
	where $r \in [0,1]$ is a memory parameter and $E_i(t)$ is obtained from the relation:
	\begin{equation}
	E_i(t)= \begin{cases}
	(p_{off}^k-C_{prod^k}) \cdot g_{off}^k & \mbox{if the bid is accepted} \\
	e \cdot m^i_k(t-1)/\left(N-1\right)  &  \mbox{otherwise},
	\end{cases}
	\end{equation}
	for all k, where $e \in [0,1]$ is an experimental parameter that assigns a different weight to played and non-played actions.
\end{enumerate}

To the best of our knowledge, Roth-Erev algorithms were previously applied to training agents based on the exclusive use of historical data with their limited relevance for current and future power distributions in electricity grids. In this paper, following the guidelines presented in \cite{MuredduPONE2015}, we overcome the usage of outdated data by performing the training on realistic system configurations, which are synthetically generated as we explain in the next section.

\section{Dataset}
For a correct representation of the market phase,  we need a detailed description of the power transmission system in space and time, in terms of network nodes, branches and generators.
The reference configuration of the power system is obtained by combining three datasets.

\begin{figure}[!t]
	\centering
	\includegraphics[width=2.5in]{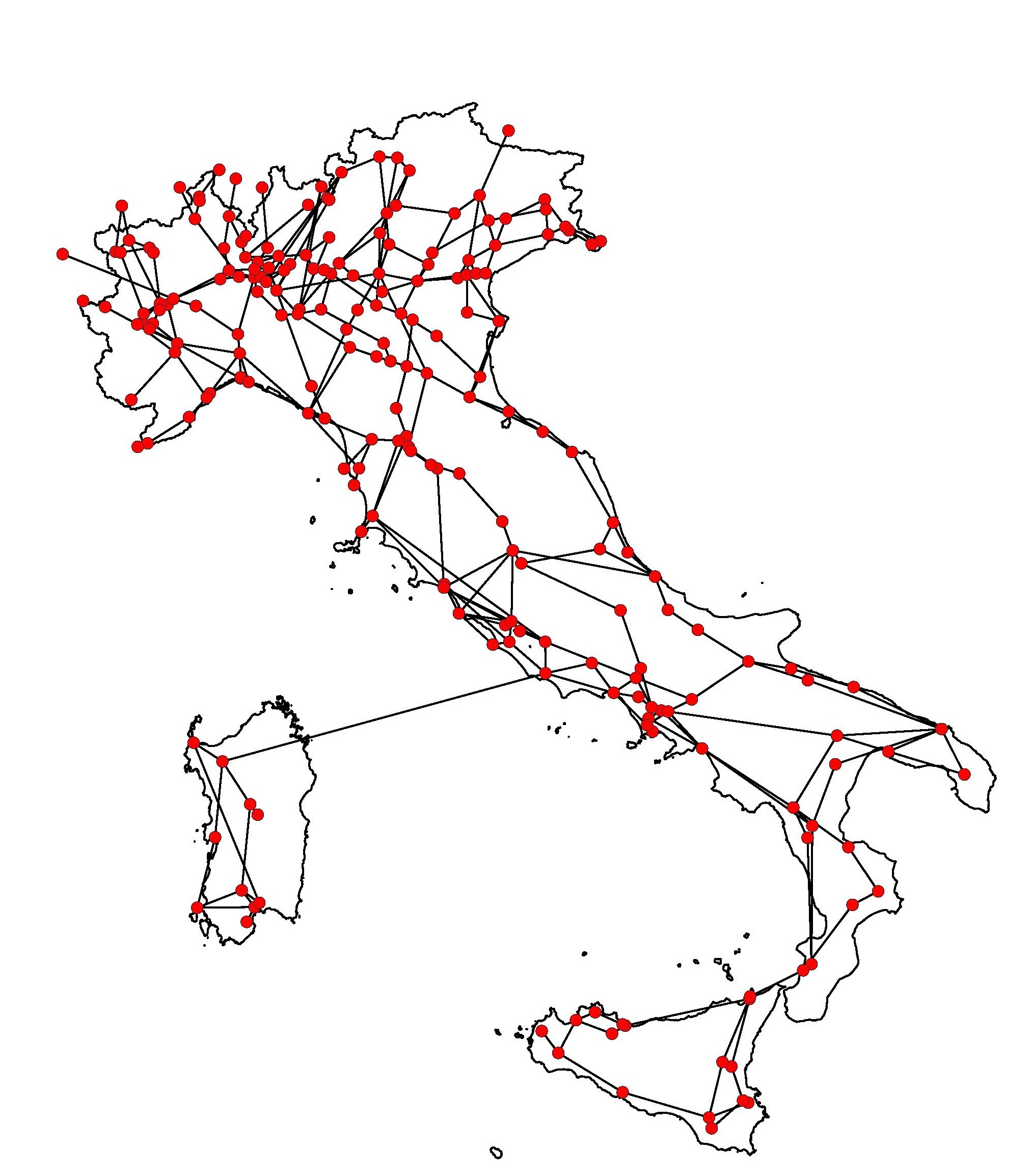}
	\caption{Geographical representation of the topology of the reference network, chosen as the 2011 Italian transmission grid with 220 and 380 KV nodes, together with connections pointing to neighboring countries.}
	\label{fig:ita}
\end{figure}

The first dataset is related to the characteristics of the power system (from the TERNA website \cite{terna}) and includes the geo-referenced position of every 220 and 380 KV substations together with their electrical characteristics, the geo-referenced position of the conventional generators, together with their power rates and power ramp limits, and the electrical characteristics of the power network. A geographical representation of these data is depicted in figure \ref{fig:ita}.

The second dataset (from the GME website \cite{gmesite}) reports the detailed time evolution of  production/consumption every 15 minutes of a reference day in the winter period 2011-2012, so that 96 data sets per day are available.

The third dataset is obtained from Atlasole and Atlavento (see the website \cite{Atlasole, Atlavento}). These sites were made available by the Italian energy services authority \cite{gmesite}. They contain the full georeferenced information on each Italian PV and wind generator, such as the installed power and technology.

Combining these datasets, we reconstructed the time evolution of power production and consumption in steps of 15 minutes over a full day in the winter period 2011-2012 for all  six market zones (virtual ones are excluded). These data ensured already a balanced grid performance, based on optimal DC-power balancing, when the conventional power production was adjusted accordingly. Moreover, starting from the real distribution of installed RES capacity, we extrapolated  the data to starting configurations with a different percentage of RES production. We separately assumed aggregated wind and photovoltaic  productions to take desired values, given by $P_{RES}(t) = P_{\%}\cdot L(t)$, with $L(t)$ being the total load that is kept fixed and $P_{\%}$ being the percentage of load covered by RES. Biomass production as a third type of RES was not considered to contribute to fluctuations, as energy production from biomass is easily controlled. We then estimated the required adapted production by conventional generation to cover the load by means of an optimal DC-power flow. As result we obtained initial configurations that lead to a stable grid performance for daily peak shares $P_{\%}$ of 24\% (which was the actual one used in \cite{GSE_RES}) up to and including 60\%, which is nowadays already temporarily achieved. For values larger than 60\% it became increasingly difficult to adjust the conventional power production to compensate for the increased amount of renewables.

\section{Results}
Based on the approach as outlined in the previous section~\ref{sec:methods}, we estimate the impact of an increasing share of wind and PV generation on the Italian balancing market. In particular, we identify a change in the daily market volumes and costs, due to an increased percentage of fluctuating production, introduced by these sources.

We tested the balancing market phase in five scenarios, characterized by a RES power production  with a share of $P_{\%}$, chosen as 24\% (as of the reference day \cite{GSE_RES}), 30\%, 40\%, 50\% and 60\% of the total load $L$. It should be noticed that the size and variation of fluctuations in the power grid are heterogeneous, as they depend on the nature of the fluctuating quantity being load, wind or photovoltaics. Their respective strengths are described by Gaussian distributions with fluctuation parameters $\sigma_l = 0.1$, $\sigma_{PV} = 0.08$ and $\sigma_{w} = 0.1$ \cite{Lu2013}, that is, chosen from real data, and in case of the assumed Weibull-distributed fluctuations of wind, chosen as $a = 2$ and $\lambda = \frac{P_w}{\Gamma(\frac{3}{2})}$ with $P_w$ the wind production from the reference configuration.

\begin{figure}
	\centering
	\includegraphics[width=\linewidth]{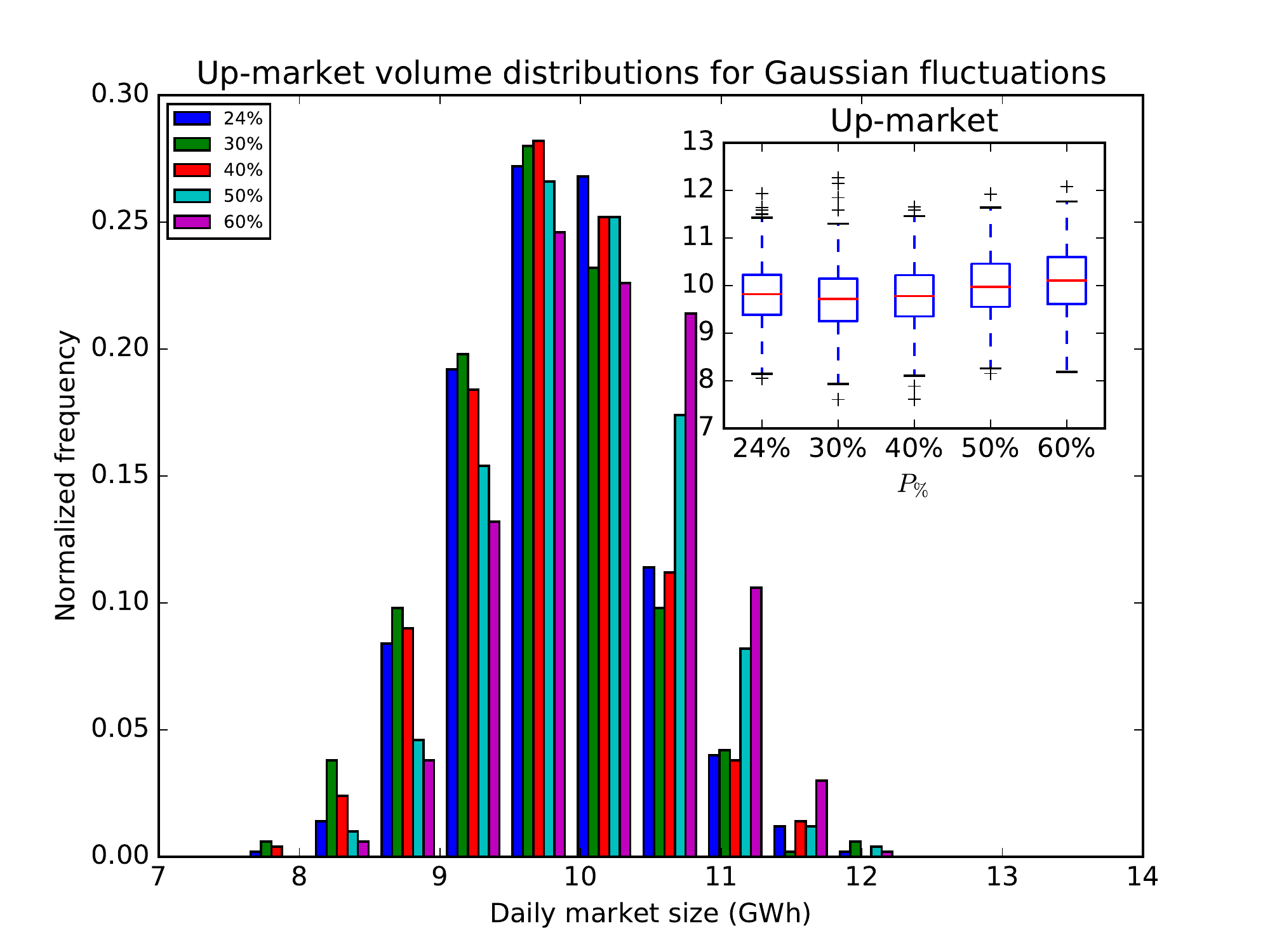}
	\caption{Distribution of daily volumes for the up-market. The results are presented as barplot showing five different values of $P_\%$. The peak value of each distribution is stable, but the distributions for larger $P_\%$-values  are more skewed to high values than they are for low percentages of RES.}
	\label{fig:marketvolumes}
\end{figure}

Next it is of much interest how these fluctuations sum up over all nodes of the grid and over a whole day. The resulting overall deviation from the total balanced power of the starting reference configuration can be positive or negative. Figure~\ref{fig:marketvolumes} shows a distribution of cases when it turns out to be negative, so that the up-energy balancing market has to compensate for this missing power. Its total volume in GWh (sum over all nodes of mismatched energy) measured over all configurations of an ensemble, over the six zones of the Italian grid, and over the reference day  yields the histogram of figure~\ref{fig:marketvolumes},

\noindent normalized over the total number of events, to predict the frequency of the various volume events. Different colors code the different scenarios, characterized by the varying contributions of RES. We present the distributions in two ways, as barplots, and as boxplots \cite{Tukey1977}. While the median (expected volume) only slightly increases for a higher amount of RES, the number of extreme market sizes  of more than 10 GWh considerably increases by roughly 50\% between 24\% and 60\% of RES. The fact that the median is relatively insensitive to the amount of RES may be due to the fact that the strongest fluctuations in size are due to load fluctuations. However, the whole distribution gets more skewed for a larger amount of RES. Therefore rare but large market sizes should lead to high energy prices in the corresponding market sessions. This must be expected even more in view of the nonlinear relation between price and demand, which is ruled by the power stack function \cite{eydeland1999fundamentals, geman2006understanding}, emerging from the market session (agent based modeling part), so that moderate increases in the demand can lead to high changes in the price. 

\begin{figure}
	\centering
	\includegraphics[width=\linewidth]{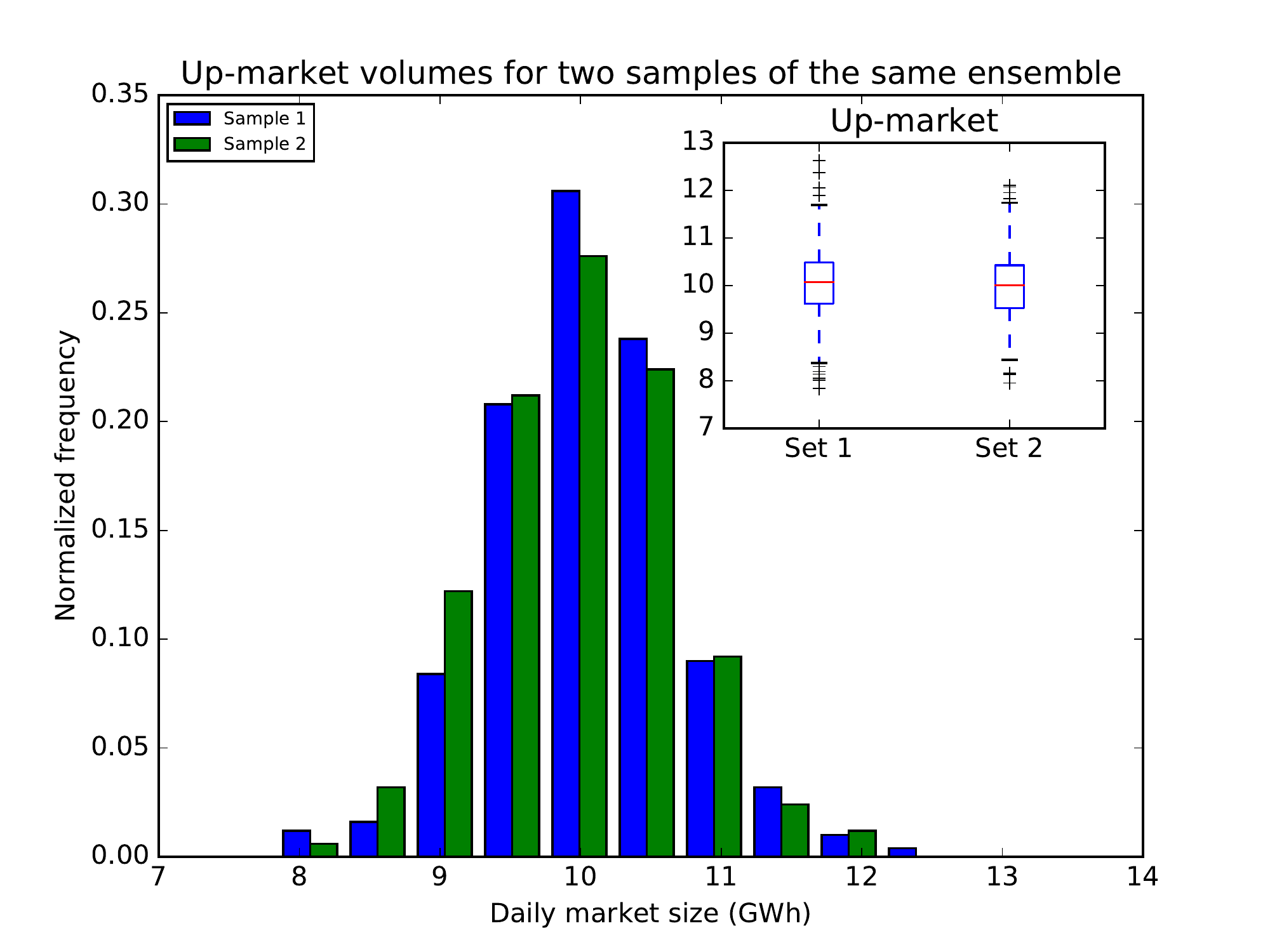}
	\caption{A comparison between the distributions of market volumes of two different samples (green and blue) for  $P_\% = 60 \%$. The plot is organized as  barplot showing the two distributions in the form of histograms, and as a boxplot in the inset. The two boxes of the boxplot show the distribution's median (red line), the first and third quartiles (blue lines next to the red one), and its statistical minimum and maximum, together with the outliers.}
	\label{fig:marketvolumescomparison}
\end{figure}

Here a remark is in order about the choice of the ensemble size, over which the observables are measured: We have chosen thousand configurations, after 3000 configurations were reserved for the learning phase of the agents.  This size seems to be sufficient, see figure~\ref{fig:marketvolumescomparison}, in which we compare the market volume for two samples of an ensemble with the same reference configuration and with 60\% of renewables (for which the contribution of fluctuations is most pronounced). Therefore the differences here are of purely statistical nature and due to the ensemble size. As it is seen from figure~\ref{fig:marketvolumescomparison}, the median and quartiles of the distributions vary then less than 1\%, differences are seen in the outliers.

\begin{figure}
	\centering
	\includegraphics[width=\linewidth]{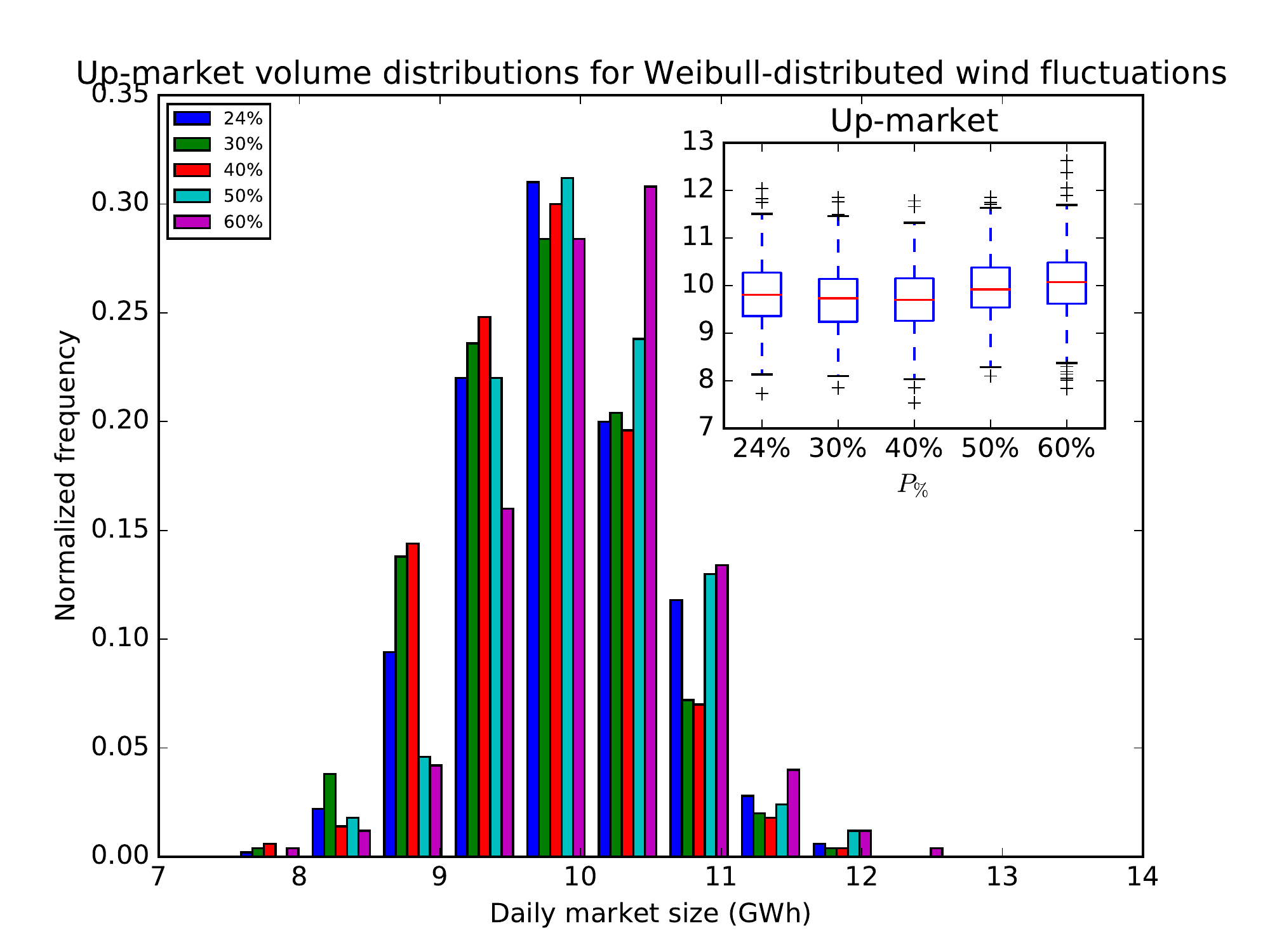}
	\caption{Distribution of daily volumes for the up-market, with wind fluctuations being Weibull-distributed, for five values of RES. Again the plot is organized as  barplot showing the five distributions in the form of histograms, and as a boxplot in the inset. The largest impact is seen in the mode of the $P_\% = 60\%$ distribution.}
	\label{fig:marketvolumeweibull}
\end{figure}

\noindent Moreover, the same analysis of the market volume was performed by assuming the wind fluctuations to be Weibull-distributed. The results are shown in figure~\ref{fig:marketvolumeweibull}.
The results for the market do not sensibly change as long as $P_\% \leq 50\%$, while for $P_\% = 60\%$  the fat tail of wind fluctuations has an impact on the volume distribution. In particular, the mode of the distribution is shifted from about 9.75 GWh to 10.5 GWh. So for larger contributions of renewables one should be aware of the possibility that the non-Gaussian fluctuations in power generation finally may shift the prices and costs towards higher values, which we here have not pursued.  

The actual results for the market prices and costs during one day under the assumption of Gaussian fluctuations are displayed in the following figures. In figure \ref{fig:costs_daily_stacked} we show the probability distribution of the market costs per day, for different values of $P_{\%}$.  Again different colors represent the different scenarios, the height of the columns is a measure for  the probability to have a daily cost in the covered interval. As shown in the boxplot in the inset, the median and quartiles of these distributions do not sensibly differ, similarly to figure~\ref{fig:marketvolumes}. This does not come as a surprise, as the balancing market is only sensitive to the power mismatch $S^j$, where the amount of RES mainly influenced the tails of the distribution rather than the median. So, also here the effect of a larger contribution of renewables is visible in the  tail of the histogram of the main figure: high values of RES generation cause more likely very high costs. For example, the number of events, for which the costs exceed 3 million Euro is roughly 4\% for $P_\% = 24\%$ as compared to 5\% for $P_\% = 60\%$ renewables.

\begin{figure}
	\centering
	\includegraphics[width=\linewidth]{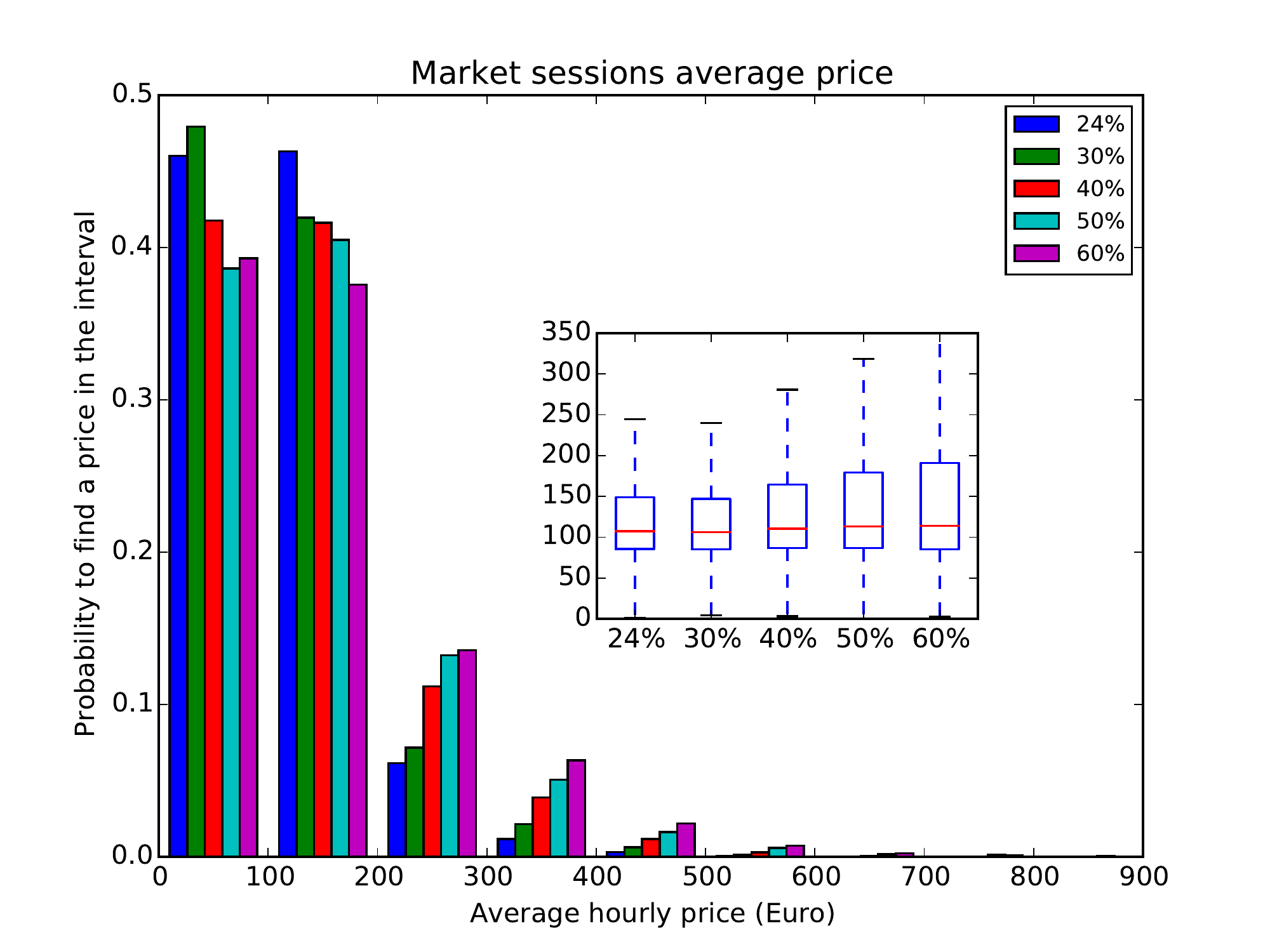}
	\caption{Probability distribution of the amount of sold energy, averaged over 4 sessions per hour, recorded  over 1000 configurations of the ensembles for each of the 24 hours per day and the six price zones,  for different values of $P_\%$. An increase in $P_{\%}$ causes an increase in the amount of energy that is sold at high prices.}
	\label{fig:Prices_histogram}
\end{figure}

Similar features are observed if the distribution of daily costs are resolved in energy prices averaged over four data per hour over all hours a day, see figure~\ref{fig:Prices_histogram}, where the third quartile and upper whiskers reflect a fat tail in the price distribution.

\begin{figure}
	\centering
	\includegraphics[width=\linewidth]{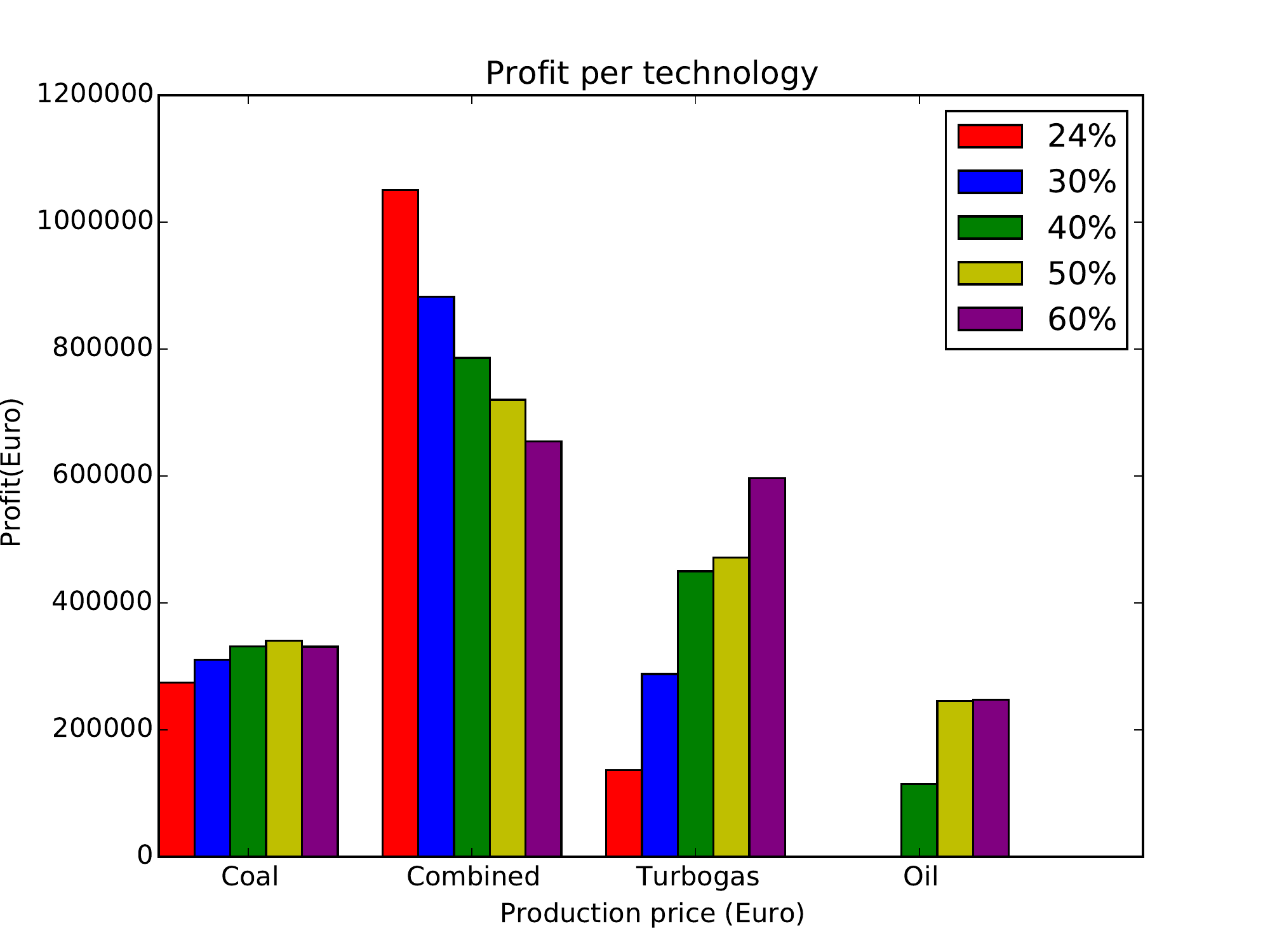}
	\caption{Average profit made by generators of different technologies, for different values of $P_{\%}$. Generators with high production costs are more likely called in the market when a high amount of renewables is present.}
	\label{fig:gen_profit}
\end{figure}

As a further result of the simulations,  figure \ref{fig:gen_profit} shows the average profit made by generators of different technologies during the day, for different values of $P_{\%}$. It was calculated from the outcomes of the market sessions for each configuration, sorted with respect to the used  technology of the generators, represented by the agents, and then averaged over the ensembles, the zones and the day. We list generators based on coal, combined cycle, turbogas, and oil technologies. It is worthwhile to stress how the average profit changes by changing $P_{\%}$. An increase in $P_{\%}$ causes an increase in the profit made by fast generators like turbogas or oil. When high market volumes become likely due to strong fluctuations,  the market authority will be forced to accept very high bids, this way encouraging a higher risk propensity for this type of generators.

\begin{figure}
	\centering
	\includegraphics[width=\linewidth]{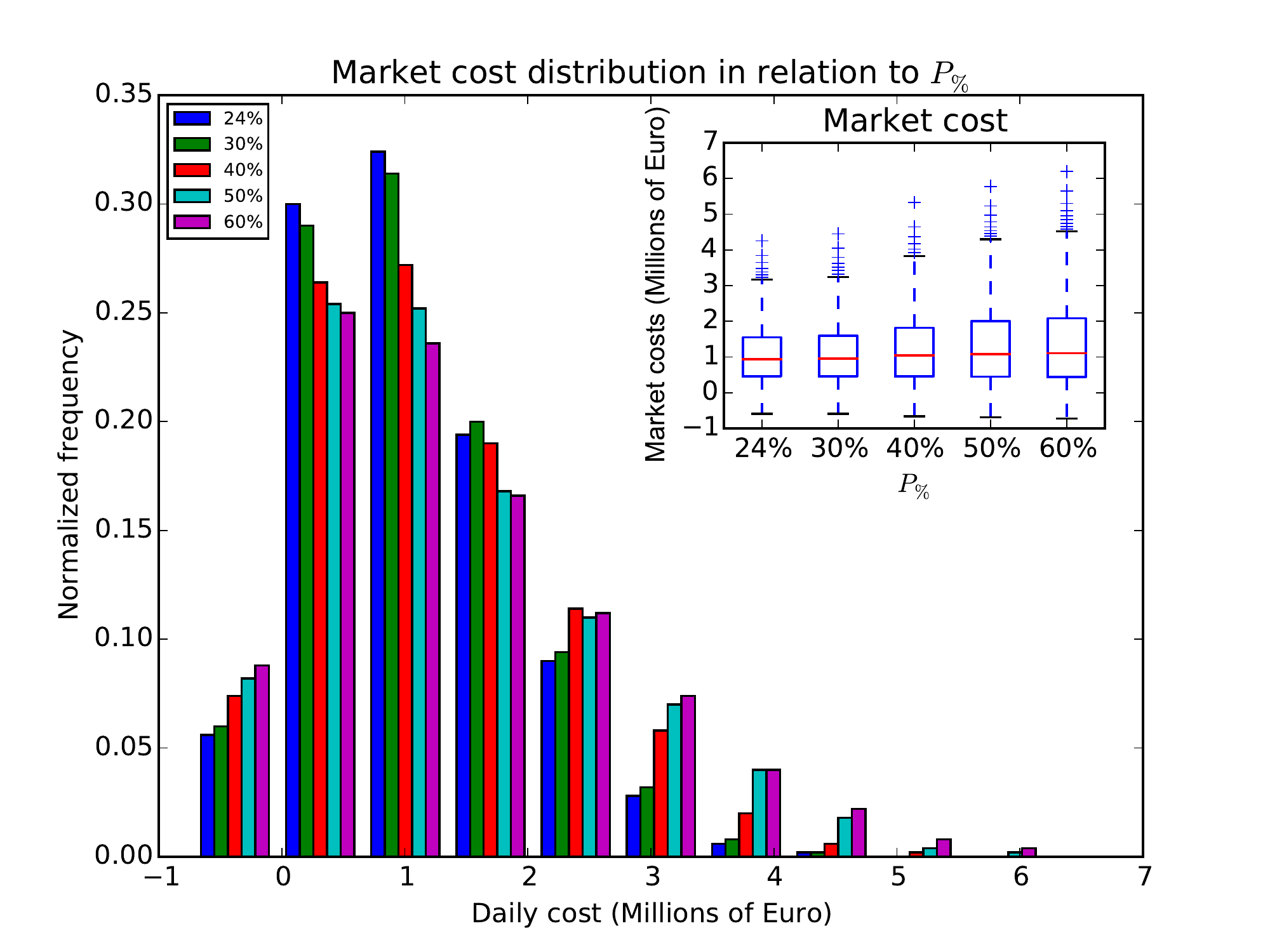}
	\caption{Predicted distribution of the daily balancing market costs, for different values of $P_{\%}$.  Although the prices' medians and quartiles do not sensibly increase with the increase in the share of RES, more events with extreme costs are observed.}
	\label{fig:costs_daily_stacked}
\end{figure}

\section{Conclusions}
Following the methodology as proposed in \cite{MuredduPONE2015}, we combined agent based modeling with a description from statistical physics by calculating the power mismatch over an ensemble of configurations that differ by small deviations from a stable power grid  configuration. The power mismatch entered as market volume in the energy balancing market, whose agents offer energy from a certain learned distribution of prices, until the missing amount of energy is covered by their offers. The original fluctuations in load and renewables lead to fluctuations in the daily costs, handled in the balancing market, and the fluctuating energy prices. While the average values of prices and actual costs only slightly increased for a higher percentage  of shared renewables, the shape of the distributions became more skewed, and the number of accepted offers at extreme prices considerably increased with an increasing amount of renewables. So far we have reduced the production of conventional generators as required by the artificially increased production of RES, while keeping the load and geographical location of production places fixed, that is, without closing any conventional production places. Moreover, power reserve in the background was kept for sale only from online conventional generators, while different technologies may share the reserve in future grid extensions. Also the underlying datasets should be further updated in view of subsequent real extensions of RES in the grid.

The methodology can be extended towards including other market phases than the short-time balancing market, different strategies of the agents in placing their bids, different price policies beyond the mere minimization of the energy-balancing market costs; also additional players may be included in the market games. So far they were restricted to agents from conventional production places. At the same time high fluctuations of RES may be damped by installing storage devices: before the fluctuations are fed into the grid, high peaks may be cut-off via storing some part of the production peaks. In any case, our methodology together with these various possible extensions allow for a quantification of the risk of high price peaks. So they aid decisions on whether countermeasures against such peaks are worth the effort, or the rare peaks are so rare to be safely ignored.

\section*{Acknowledgment}
The authors gratefully acknowledge the support from the German Federal Ministry of Education and Research (BMBF grant no. 03SF0472D (NET-538-167)).

\end{document}